\documentclass[aps,prd,reprint,nofootinbib,superscriptaddress]{revtex4-2}

\usepackage{amsmath,amssymb,bm}
\usepackage{graphicx}
\usepackage{xcolor}
\usepackage{hyperref}
\usepackage{orcidlink}

\newcommand{\dd}{\mathrm{d}}
\newcommand{\ket}[1]{\left|#1\right\rangle}
\newcommand{\bra}[1]{\left\langle#1\right|}

\begin{document}

\title{Superconducting Qubits as Quantum Targets for Direct Particle Scattering}

\author{Doojin Kim\,\orcidlink{0000-0002-4186-4265}}
\email{doojin.kim@usd.edu}
\affiliation{Department of Physics, University of South Dakota, Vermillion, South Dakota 57069, USA}
\author{Kyoungchul Kong\,\orcidlink{0000-0003-4515-7303}}
\email{kckong@ku.edu}
\affiliation{Department of Physics and Astronomy, University of Kansas, Lawrence, Kansas 66045, USA}
\author{Myeonghun Park\,\orcidlink{0000-0002-4010-7860}}
\email{parc.seoultech@seoultech.ac.kr}
\affiliation{Institute of Convergent Fundamental Studies, Seoultech, Seoul 01811, South Korea}
\affiliation{School of Natural Sciences, Seoultech, Seoul 01811, South Korea}

\begin{abstract}
We introduce a novel approach to particle detection in which a superconducting qubit itself serves as the scattering target and a resolved transition between circuit eigenstates constitutes the observable final state. The scattering amplitude factorizes into the incident-particle current and a finite-momentum electromagnetic transition response of the qubit. This momentum dependence is essential in that the energy transfer is fixed by the microwave-scale level spacing, whereas a nonrelativistic incident particle can transfer a spatial momentum parametrically larger than that energy and thereby resolve the device geometry. For charge-dominated processes, we recast the response as an induced-offset-charge kernel obtained from an auxiliary electrostatic weighting potential, providing a practical route from circuit geometry to a differential scattering cross section. We illustrate the framework with an Xmon-style transmon and photon-mediated scattering of dark matter. The exceptionally small qubit transition threshold opens sensitivity to an eV-to-keV mass regime that is largely unexplored by existing direct-detection experiments.
\end{abstract}

\maketitle

\noindent\textbf{Introduction.---}
Quantum sensors offer access to energy deposits far below the thresholds of conventional detector technologies~\cite{Essig:2022dfa,Cecil:2022fiv,Buchmueller:2022djy}. Superconducting qubit circuits, in particular, have received rising attention as single-microwave-photon counters, coherent-field sensors, and monitors of quasiparticle production and other weak energy-deposition channels~\cite{Dixit:2020ymh,Chen:2022quj,Das:2022srn,Fink:2023tvb,Das:2024jdz,Ramanathan:2024hsf,Linehan:2025suv,Li:2025hqk,Hochberg:2026mdl,Dong:2025mdk}. Other quantum platforms seek microscopic excitations created by particle interactions in bulk materials or trapped systems~\cite{Budker:2021quh,Mei:2025fxj,Mei:2025ozs}. These developments exploit quantum devices as exceptionally low-threshold readout elements, but they do not require the internal transition of a superconducting qubit itself to be the elementary particle-scattering process.

Direct scattering at similarly small energy scales has also been studied in condensed-matter and trapped-particle targets. In superconductors and other electronic materials, the target is commonly described through microscopic many-body response functions or dielectric screening~\cite{Hochberg:2021pkt,Knapen:2021run,Knapen:2021bwg,Lasenby:2021wsc}; graphene Josephson-junction concepts use the junction as a bolometric sensor of energy deposited in the graphene sheet itself~\cite{Kim:2020coy,Jeong:2026buu}; and trapped-ion proposals register motional excitations caused by a passing particle~\cite{Budker:2021quh}, with recent work extending coherent free-electron interactions to trapped-ion processors and more general quantum two-level systems~\cite{Pescoller:2026zhp,Grzesik:2026yyg}. The construction in this Letter addresses a distinct target degree of freedom: the initial and final states are eigenstates of the superconducting circuit itself. Consequently, the relevant target quantity is not merely the total deposited energy, but the matrix element of the circuit electromagnetic current between two resolved quantum states at the momentum carried by the exchanged photon.

\begin{figure}[b]
    \centering
    \includegraphics[width=0.35\textwidth]{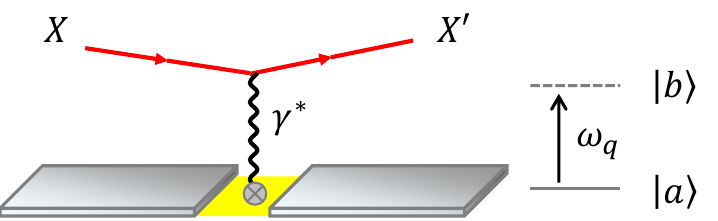}
    \caption{Schematic diagram of direct transition scattering on a superconducting circuit. An incident particle transfers \(q^\mu=(\omega_q,\vec q)\) through an off-shell photon and promotes the circuit from \(\ket a\) to \(\ket b\).}
    \label{fig:schematic}
\end{figure}

In this context, we consider an incident particle $X$ that scatters off to an outgoing particle $X'$\footnote{$X'$ may or may not be the same as the incident species $X$.} via an off-shell photon exchange with a superconducting circuit, promoting the circuit between two stationary eigenstates, $|a\rangle$ and $|b\rangle$ with eigenenergies $E_a$ and $E_b$, respectively (see also FIG.~\ref{fig:schematic}):
\begin{equation}
X(p)+\ket a\to X'(p')+\ket b,
\qquad \omega_q\equiv E_b-E_a>0.
\label{eq:process}
\end{equation}
The associated energy-momentum transfer is defined as
\begin{equation}
q^\mu\equiv p^\mu-p'^\mu=(\omega_q,\vec q),
\qquad Q\equiv |\vec q|,
\label{eq:qdef}
\end{equation}
where the energy transfer $E-E'$ is also set by the characteristic (but tunable) circuit excitation $\omega_q$.

We emphasize that the chip and its support may absorb the spatial recoil at negligible energy cost, while the resolved circuit excitation fixes the deposited energy to \(\omega_q\). 
In resonant absorption of an on-shell microwave photon one has \(Q=\omega_q\), whereas for a nonrelativistic particle with speed $v$, the momentum transfer is generically of order \(\omega_q/v\) or larger. Thus typical GHz-scale transitions do not imply spatially long-wavelength probes, i.e., the scattering particle can resolve the finite structure of the capacitor, ground plane, and surrounding dielectric environment.

This setup also differs conceptually from a coherent-field measurement. A classical or highly occupied bosonic field can drive a qubit coherently, allowing a transition amplitude or phase to build during an interrogation interval~\cite{Chen:2023swh}. By contrast, the incoming particle here is an asymptotic scattering state; a single event transfers definite energy and momentum, leaving the particle in a different outgoing state. The observable is therefore an inelastic scattering rate, with the circuit transition serving as the tag of the microscopic interaction. This distinction makes it possible to use familiar scattering-theory language while retaining the genuinely quantum internal structure of the target.

The resulting problem is naturally phrased in terms of a finite-momentum transition response rather than a dipole matrix element alone. Below we first construct a general current-based description, then specialize it to the charge response of a transmon, taking an Xmon-style device as a concrete realistic example. We finally use superlight dark matter as a representative application. Detailed derivations, alternative geometries, multi-qubit collective states, and additional detector configurations are presented in the companion study~\cite{Kim:2026companion}, while we focus on establishing the general framework in this Letter. 

\medskip

\noindent\textbf{Formalism.---}
At leading order in the electromagnetic interaction, the transition amplitude can be written as
\begin{equation}
 i\mathcal M_{ba}
 =j_X^\mu(p',p)\,[-iD_{\mu\nu}(q)]\,
 \mathcal J_{ba}^\nu(\vec q),
\label{eq:masteramp}
\end{equation}
where \(j_X^\mu\) is the electromagnetic current of the incident particle $X$ and \(D_{\mu\nu}\) is the photon propagator or, in a material environment, the appropriate dressed electromagnetic Green function. All information about the circuit transition is encapsulated in
\begin{equation}
\mathcal J_{ba}^\mu(\vec q)
\equiv
\bra b\int\dd^3x\,e^{i\vec q\cdot\vec x}
\hat J_Q^\mu(\vec x)\ket a
=\bigl(\rho_{ba}(\vec q),\vec j_{ba}(\vec q)\bigr).
\label{eq:Jtransition}
\end{equation}
For stationary states, current conservation gives $q_\mu\mathcal J_{ba}^\mu=0$, or equivalently, 
\begin{equation}
\omega_q\rho_{ba}=\vec q\cdot\vec j_{ba}.
\label{eq:continuity}
\end{equation}
The charge density therefore fixes the longitudinal electric response, while transverse currents contain additional magnetic and transverse-electric information. 
For charge-dominated transitions, a useful microscopic characterization of the circuit response is provided by 
\begin{equation}
\rho_{ba}(\vec q)=\bra b\int \dd^3x\, e^{i\vec q\cdot \vec x}\hat\rho_Q(\vec x)\ket a.
\label{eq:QTFF}
\end{equation}
For \(b\neq a\), Eq.~\eqref{eq:continuity} implies \(\rho_{ba}(0)=0\). The first nonzero term in a formal long-wavelength expansion can therefore be dipolar, but such an expansion need not converge rapidly when \(QL\gtrsim1\), with \(L\) a characteristic device dimension. We consequently retain the full \(\vec q\) dependence. The same dependence also keeps track of the orientation of the momentum transfer relative to an anisotropic (planar) circuit. Thus a finite-size qubit can, in principle, carry directional information through its transition probability although the final internal state itself has no spatial direction label (see e.g., Refs.~\cite{Boyd:2022tcn,Hochberg:2021ymx,Jeong:2026buu} for the directional measurements in similar contexts). For the benchmark below, we average over device orientation and leave directional operation to future work.

Treating the qubit as a localized fixed target, with relativistically normalized incident and outgoing particles, we find that the differential scattering cross section is given by
\begin{equation}
\frac{\dd\sigma_{a\to b}}{\dd Q^2}
=\frac{1}{16\pi p^2}
\overline{|\mathcal M_{ba}(Q)|^2}.
\label{eq:dsigdq_general}
\end{equation}
The fixed-target normalization is appropriate because the translational recoil energy of the macroscopic support is negligible on the scale of the internal transition, while momentum conservation is maintained by the full chip-plus-support system. The qubit itself therefore need not be assigned a center-of-mass recoil energy. This is the same separation implicit in Eq.~\eqref{eq:process}; the discrete internal energy is measured, whereas the spatial momentum probes the extended circuit. 

The discussion so far is independent of the qubit architecture and of the identity of $X$. Choosing a device and a particle model amounts to specifying the two currents.
For a transmon, the Hamiltonian operator $\widehat H_T$~\cite{Koch:2007hay,Krantz:2019jkw,Blais:2020wjs} is
\begin{equation}
\widehat H_T=4E_C\hat n^2-E_J\cos\hat\varphi,
\label{eq:transmonH}
\end{equation}
where $E_{C(J)}$ is the charging (Josephson) energy and \(\hat n=\hat Q/(2e)\) and $\hat\varphi$ are the dimensionless charge-number operator and superconducting phase operator, respectively. In a single collective charge mode, the transition density may be represented as
\begin{equation}
\hat\rho_Q(\vec x)=2e\,\hat n\,\eta_\rho(\vec x),
\qquad \int\dd^3x\,\eta_\rho(\vec x)=0,
\label{eq:chargeprofile}
\end{equation}
where $\eta_\rho$ describes the spatial profile associated with $\hat n$. We therefore identify
\begin{equation}
\rho_{ba}(\vec q)=2e\,n_{ba}f_\rho(\vec q), \qquad
f_\rho(\vec q)\equiv\int\dd^3x\,e^{i\vec q\cdot\vec x}\eta_\rho(\vec x),
\label{eq:factorization}
\end{equation}
with $n_{ba}\equiv\bra b\hat n\ket a$. 
This decomposition implies that \(n_{ba}\) carries the internal transition strength and selection rule, whereas \(f_\rho\) describes momentum matching to the spatial charge pattern. For the \(0\to1\) transition in the usual transmon regime \(E_J\gg E_C\),
\begin{equation}
\omega_q\simeq\sqrt{8E_JE_C}-E_C,
\qquad
|n_{10}|^2\simeq\frac{\omega_q+E_C}{16E_C}.
\label{eq:transmon_matrix}
\end{equation}

For a fabricated planar circuit, however, separating a vacuum propagator from a prescribed charge profile is not the most useful representation. The same external charge that couples to the island also polarizes the surrounding ground plane and dielectric interfaces. We therefore encode the complete quasi-static electric response in the offset charge induced on the transmon mode. Consider an auxiliary electrostatic problem in which the island is assigned unit potential and the surrounding grounded conductors zero potential. The dimensionless weighting potential \(u_Q(\vec x)\) obeys the generalized Laplace equation:
\begin{equation}
\vec\nabla\!\cdot[\epsilon(\vec x)\vec\nabla u_Q(\vec x)]=0,
\label{eq:weighting}
\end{equation}
where $\epsilon$ is the usual electric permittivity.
The unit-potential condition is only a normalization of this auxiliary problem, not the physical operating voltage of the qubit. 
Using the standard weighting-potential construction underlying the Shockley-Ramo theorem~\cite{Shockley:1938itm,Ramo:1939vr}, a point charge \(q_X\) at \(\vec x\) induces the mode charge \(Q_{\rm ind}=-q_X u_Q(\vec x)\).
Defining the dimensionless response function $\kappa_Q(\vec x)$ as \(\kappa_Q(\vec x)\equiv-u_Q(\vec x)\), an arbitrary quantum charge source produces
\begin{equation}
\delta\hat n_g
=\frac{1}{2e}\int\dd^3x\,
\kappa_Q(\vec x)\hat\rho_X(\vec x).
\label{eq:dng}
\end{equation}
where $\hat \rho_X$ denotes the charge-density operator of the incident particle,
\begin{equation}
    \hat\rho_X(\vec x)
    =
    q_X\,
    \delta^{(3)}
    \left(
        \vec x-\hat{\vec x}_X
    \right). \label{eq:dng_rho_def}
\end{equation}
Expanding Eq.~\eqref{eq:transmonH} to first order in this induced offset charge yields
\begin{equation}
\widehat H_{\rm int}
=-\frac{4E_C}{e}\hat n
\int\dd^3x\,\kappa_Q(\vec x)\hat\rho_X(\vec x).
\label{eq:Hint}
\end{equation}
Defining the response function in momentum space
\begin{equation}
\widetilde\kappa_Q(\vec q)
\equiv\int\dd^3x\,e^{i\vec q\cdot\vec x}\kappa_Q(\vec x),
\label{eq:kappatilde}
\end{equation}
with 
\(
    i\mathcal M_{ba}= -i\bra{p',b}\hat H_{\rm int}\ket{p,a},
\)
we obtain
\begin{equation}
\mathcal M_{ba}
=\frac{4E_C}{e}\,n_{ba}\,j_X^0(p',p)\,
\widetilde\kappa_Q(\vec q),
\label{eq:Mdevice}
\end{equation}
and hence
\begin{equation}
\frac{\dd\sigma_{a\to b}}{\dd Q^2}
=\frac{E_C^2|n_{ba}|^2}{\pi e^2p^2}
\overline{|j_X^0|^2}\,\langle|\widetilde\kappa_Q(\vec q)|^2\rangle,
\label{eq:master_xsec_device}
\end{equation}
where $\langle ... \rangle$ implies the device-orientation-averaged quantity.

This expression suggests that \(\widetilde\kappa_Q\) can be interpreted as a device-level finite-momentum response; it incorporates conductor geometry, dielectric screening, and electrostatic boundary conditions in the same object that enters the particle-scattering cross section. Operationally, the weighting-potential construction is useful because one electrostatic solution determines the induced charge from a source placed anywhere in space; quantization of that source then converts the same kernel directly into a scattering matrix element. In an idealized vacuum geometry, this description can be rearranged into a photon propagator multiplying a transition charge form factor, whereas for a fabricated circuit, it automatically includes the electrostatic redistribution of charge on nearby conductors. 
The quasi-static treatment is well suited to microwave transitions in a sub-millimeter circuit, for which the electromagnetic wavelength associated with \(\omega_q\) is much larger than the device. Eq.~\eqref{eq:masteramp} nevertheless remains the more general starting point when transverse or retarded electromagnetic response is important.

The transition-current and weighting-potential descriptions should be viewed as two representations of the same linear electromagnetic coupling at different levels of coarse graining. A microscopic material-response calculation (e.g., Refs.~\cite{Hochberg:2021pkt, Knapen:2021run, Knapen:2021bwg, Lasenby:2021wsc, Kim:2020coy, Jeong:2026buu}) retains momentum- and frequency-dependent polarization inside a translationally invariant medium, whereas the device treatment used here retains the finite conductor boundaries and replaces the local material response by macroscopic dielectric parameters. This distinction matters for a lithographically patterned qubit, where ground planes and electrode edges can redistribute the induced charge over distances comparable to the inverse momentum transfer. Conversely, if microscopic quasiparticle, dielectric, or transverse-current dynamics become relevant, they can be restored by replacing the quasi-static kernel with the corresponding dressed Green function in Eq.~\eqref{eq:masteramp}. The framework therefore does not rely on a particular electrostatic model; the Xmon calculation below is a concrete realization of the general transition response.

For the numerical benchmark, we use an Xmon-style planar transmon based on the capacitor dimensions specified in Ref.~\cite{Barends:2013zvp}: a cross-shaped island with arm length \(L=160~\mu\mathrm m\), strip width \(S=24~\mu\mathrm m\), and a \(W=24~\mu\mathrm m\) gap to the coplanar ground. We solve Eq.~\eqref{eq:weighting} in a planar effective-interface treatment with effective permittivity \(\epsilon_{\rm eff}=5.2\), using an approximately \(1~\mathrm{mm}\times1~\mathrm{mm}\) computational region at \(1~\mu\mathrm m\) spatial resolution. 
The induced conductor charge distribution determines \(\widetilde\kappa_Q(\vec q)\), while the total island charge for unit potential gives \(C_\Sigma\simeq70~\mathrm{fF}\), corresponding to \(E_C/(2\pi)\simeq275~\mathrm{MHz}\). 
The numerical response is retained with its angular dependence and later orientation-averaged for the sensitivity estimates. With this pair of domain and grid, we conduct numerical calculation to estimate the response over the kinematically allowed $Q$ range.
Control lines, readout resonators, the Josephson-junction microgeometry, and package structures are omitted from this minimal Xmon realization. More detailed discussions including the construction, convergence tests, and the sensitivity of the response to such modeling choices are given in Ref.~\cite{Kim:2026companion}.

\medskip

\noindent\textbf{Application.---}
We now use superlight dark matter as a representative projectile to illustrate the kinematic regime in which the finite-momentum qubit response becomes essential.
More specifically, we consider a (nonthermally produced) complex scalar dark-matter particle \(\chi\) with effective electromagnetic current
\begin{equation}
j_\chi^\mu=g_\chi(p+p')^\mu,
\label{eq:dmcurrent}
\end{equation}
where $g_\chi$ parameterizes the coupling strength of $\chi$ to the photon.
In the nonrelativistic limit, Eqs.~\eqref{eq:transmon_matrix} and \eqref{eq:master_xsec_device} reduce to
\begin{equation}
\frac{\dd\sigma_{\chi,0\to1}}{\dd Q^2}
=\frac{g_\chi^2E_C(\omega_q+E_C)}{4\pi e^2v^2}
\langle |\widetilde\kappa_Q(\vec q)|^2 \rangle.
\label{eq:dmxsec}
\end{equation}
The transition kinematics suggests
\begin{equation}
\omega_q=Qv\cos\theta-\frac{Q^2}{2m_\chi},
\qquad
v_{\min}(Q)=\frac{\omega_q}{Q}+\frac{Q}{2m_\chi},
\label{eq:kinematics}
\end{equation}
where $\theta$ is the angle between $\vec q$ and the dark-matter velocity vector $\vec v$. At fixed speed, we find the allowed momentum-transfer limits,
\begin{equation}
Q_\pm=m_\chi v\pm
\sqrt{m_\chi^2v^2-2m_\chi\omega_q}.
\label{eq:Qpm}
\end{equation}
A kinematically allowed transition requires \(m_\chi v^2\ge 2\omega_q\). Independently, \(Q_-\ge\omega_q/v\), making explicit why the low energy threshold and the finite-momentum device response are separate ingredients; decreasing \(\omega_q\) lowers the accessible particle mass, while the remaining momentum transfer determines which spatial Fourier components of the qubit are sampled. 
For the \(5~\mathrm{GHz}\) benchmark taken for our sensitivity study, \(\omega_q\simeq20.7~\mu\mathrm{eV}\). The characteristic threshold is \(m_\chi^{\min}=2\omega_q/v^2\), which is tens of eV for a canonical halo speed \(v\sim10^{-3}\) and falls to the few-eV scale in the high-speed portion of the halo. At the same time, \(Q_-^{-1}\lesssim v/\omega_q\) is of order ten microns for \(v\sim10^{-3}\). Consequently, scattering on a conventional transmon generally resolves its finite spatial structure, making momentum-dependent device responses essential.

For a normalized laboratory velocity distribution \(f_{\rm lab}\), the rate per qubit is
\begin{align}
R_\chi={}&\frac{\rho_\chi}{m_\chi}
\int\dd^3v\,f_{\rm lab}(\vec v)\,v\,
\Theta\!\left(\frac12m_\chi v^2-\omega_q\right)\nonumber\\
&\times\int_{Q_-^2}^{Q_+^2}\dd Q^2\,
\frac{\dd\sigma_{\chi,0\to1}}{\dd Q^2},
\label{eq:rate}
\end{align}
where $\rho_\chi$ is the local dark-matter energy density.
We remark that Eq.~\eqref{eq:rate} is evaluated in the dilute-particle limit, neglecting Bose stimulation into occupied final dark-matter modes. 
For sufficiently light bosonic dark matter, the large phase-space occupation can enhance the scattering rate, so that the projected sensitivity shown shortly is interpreted to be conservative, especially toward the lowest-mass region. 
A full treatment is deferred to the companion study~\cite{Kim:2026companion}.
To evaluate $f_{\rm lab}$, we adopt a truncated Maxwell-Boltzmann halo~\cite{Lewin:1995rx} defined in the Galactic coordinate and transform it to the Earth coordinate, with \(v_0=220~\mathrm{km/s}\), \(v_E=232~\mathrm{km/s}\), \(v_{\rm esc}=544~\mathrm{km/s}\)~\cite{Lewin:1995rx,Smith:2006ym}, and \(\rho_\chi=0.4~\mathrm{GeV/cm^3}\)~\cite{Catena:2009mf,Read:2014qva}. To facilitate comparison with conventional electron-scattering searches, the coupling value determining the sensitivity is translated to the commonly-used reference cross section
\begin{equation}
\bar\sigma_e
=\frac{16\pi\alpha_{\rm em}^2g_\chi^2}{e^2q_{\rm ref}^4}\,
\mu_{\chi e}^2,
\qquad q_{\rm ref}=\alpha_{\rm em}m_e,
\label{eq:sigmabar}
\end{equation}
where $\alpha_{\rm em}$, $\mu_{\chi e}$, and $q_{\rm ref}$ are the usual electromagnetic fine structure constant, the reduced mass of the $\chi$-$e$ system, and the reference momentum transfer, respectively. 

For an array of \(N\) independently initialized qubits, exciting different qubits produces mutually orthogonal one-excitation final states. The inclusive probability therefore adds incoherently over those final states, giving \(R_N=NR_\chi\), even if the incident wave is spatially coherent across more than one qubit. 
We hence characterize the detector by the live exposure \(\mathcal E=NT_{\rm live}\). This interpretation also clarifies the role of the measurement cycle. In the rare-event limit, a window of duration \(\tau\) has scattering probability \(P\simeq R_\chi\tau\); summing many initialization-interrogation-readout cycles gives \(N_{\rm sig}=NR_\chi T_{\rm live}\), independent of how the same live time is partitioned. Finite relaxation, readout efficiency, and dead time modify the detection efficiency or effective live exposure, rather than the microscopic scattering rate. This behavior is distinct from a coherent-field search, for which amplitudes may accumulate during an interrogation interval. Collective-state enhancements likewise require a different initial many-body state and are outside the scope of this Letter; see Ref.~\cite{Dong:2025mdk} for Dicke-state enhancement in qubit sensing and Ref.~\cite{Kim:2026companion} for its finite-momentum treatment in particle scattering.

\begin{figure}[t]
    \centering
    \includegraphics[width=0.48\textwidth]{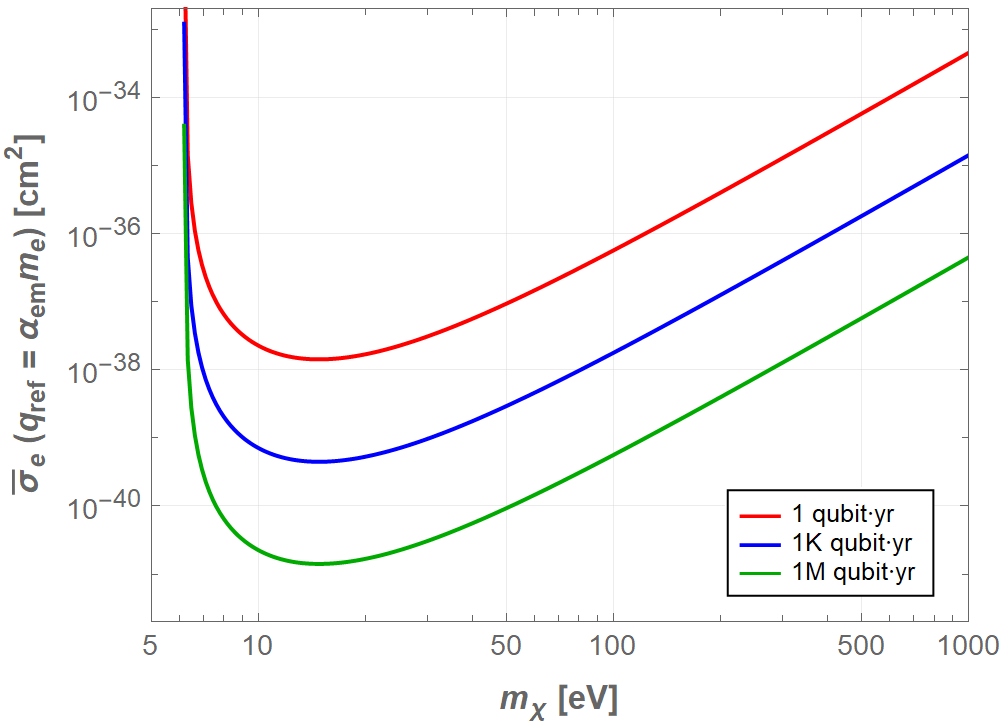}
    \caption{Projected 90\% C.L. sensitivity to \(\bar\sigma_e\) for a \(5~\mathrm{GHz}\) Xmon-style transmon. Red, blue, and green curves correspond to exposures of \(1\), \(10^3\), and \(10^6~\mathrm{qubit}\cdot{\rm yr}\), respectively, with the residual dark-count benchmark specified in the text.}
    \label{fig:sensitivity}
\end{figure}

We display the resulting reach for a \(5~\mathrm{GHz}\) Xmon-style transmon using the numerical response described above, in FIG.~\ref{fig:sensitivity}. We consider exposures of \(1\), \(10^3\), and \(10^6~\mathrm{qubit}\cdot{\rm yr}\).
These three exposures span a single-device benchmark, an intermediate scale comparable to already demonstrated thousand-qubit superconducting systems~\cite{Castelvecchi:2023ibm}, and a long-term scale motivated by million-qubit fault-tolerant architectures~\cite{Saadatmand:2024erv}.
In addition, we adopt an effective residual dark-count rate \(b_{\rm dark}=10^4~\mathrm{events/(qubit}\cdot{\rm yr)}\)~\cite{Bratrud:2025}. This value is motivated as an order-of-magnitude detector benchmark by measured sub-mHz charge-jump rates in shielded superconducting-qubit systems~\cite{Bratrud:2025}; it should not be interpreted as a microscopic model of false \(0\to1\) excitations. For an exposure \(\mathcal E\), the expected background is \(B=b_{\rm dark}\mathcal E\), which we treat as known when constructing the projected 90\% C.L. upper limit. Readout inefficiency, finite qubit relaxation time, and reset or acquisition dead time can be folded into an effective live exposure without changing the underlying scattering rate. 
In this sense, the projection separates the microscopic quantity calculated here---the probability for the particle to create a resolved circuit excitation---from the experiment-specific probability for that excitation to survive, be read out, and pass event-selection criteria. This modular separation is useful because improvements in qubit lifetime or readout fidelity change the detection efficiency without requiring a new particle-scattering calculation, while changes in electrode geometry or circuit frequency enter directly through \(\widetilde\kappa_Q\) and \(n_{10}\).

The lower edge of the projected mass reach is set by the kinetic energy needed to produce a \(5~\mathrm{GHz}\) excitation and is populated only by the fastest halo particles near threshold. The combination of a microwave-scale gap and a finite-size circuit response consequently accesses a mass range far below that of standard electron-recoil searches. 
For standard-halo dark matter without an additional boosting mechanism, existing electron-recoil searches lose sensitivity well above the mass regime considered here. Superconducting-nanowire searches currently reach masses of order tens of keV~\cite{QROCODILE:2024nqm}, while solar-reflected dark matter has extended xenon-based sensitivity to \(4.6~{\rm keV}\)~\cite{XENON:2025dpl}; graphene Josephson-junction proposals extend toward the sub-keV regime~\cite{Kim:2020coy}. 
Figure~\ref{fig:sensitivity} illustrates how the exceptionally low excitation threshold of a qubit opens sensitivity to a qualitatively new kinematic regime when the qubit transition itself is the detected scattering final state. 

\medskip

\noindent\textbf{Conclusion.---}
We have developed a framework for direct particle scattering in which a resolved transition between superconducting-qubit eigenstates constitutes the observable final state. Because the momentum transferred by a nonrelativistic incident particle need not be small compared with the inverse device size even when the deposited energy is only in the GHz range, the scattering process probes the finite-momentum electromagnetic response of the qubit rather than only its usual long-wavelength circuit matrix elements. Using an Xmon-style transmon as a representative realization, we showed how this response can be incorporated into a differential scattering cross section and illustrated the resulting sensitivity to dark matter in a mass range that is largely inaccessible to conventional direct-detection experiments. More generally, the same formalism applies to other particle species and interaction structures through the corresponding current, establishing superconducting qubits as finite-size quantum targets for particle scattering.


\acknowledgments
The work of DK is supported in part by NSF Grant No.~PHY-2609913. The work of KK is supported in part by NSF Grant No.~PHY-2609759. The work of MP is supported by the National Research Foundation of Korea (NRF) grant funded by the Korea government (MSIT) (No.~RS-2025-00564488).

\bibliography{refs}

\end{document}